\documentclass[lettersize,journal]{IEEEtran}
\usepackage{amsmath,amsfonts}
\usepackage{booktabs} 
\usepackage{caption}  
\usepackage{algorithmic}
\usepackage{algorithm}
\usepackage{array}
\usepackage[caption=false,font=normalsize,labelfont=sf,textfont=sf]{subfig}
\usepackage{textcomp}
\usepackage{stfloats}
\usepackage{url}
\usepackage{verbatim}
\usepackage{graphicx}
\usepackage{cite}
\usepackage[colorlinks,
            linkcolor=black,
            anchorcolor=black,
            citecolor=black]{hyperref}
\begin{document}

\title{M3SD: Multi-modal, Multi-scenario and Multi-language \\ Speaker Diarization Dataset}

\author{Shilong Wu
}



\maketitle

\begin{abstract}
In the field of speaker diarization, the development of technology is constrained by two problems: insufficient data resources and poor generalization ability of deep learning models. To address these two problems, firstly, we propose an automated method for constructing speaker diarization datasets, which generates more accurate pseudo-labels for massive data through the combination of audio and video. Relying on this method, we have released Multi-modal, Multi-scenario and Multi-language Speaker Diarization (M3SD) datasets. This dataset is derived from real network videos and is highly diverse. Our dataset and code have been open-sourced at \url{https://huggingface.co/spaces/OldDragon/m3sd}.
\end{abstract}

\begin{IEEEkeywords}
Speaker diarization, dataset construction, pseudo label, model fine-tuning.
\end{IEEEkeywords}

\section{Introduction}
\IEEEPARstart{S}{peaker} diarization is a significant front-end processing technology, which can solve the problem of "who spoke when" \cite{dia_review}. In interviews, broadcast news, debate competitions, conferences and other scenes, audio or video recording often involves multiple participants, which makes the speaker diarization technology have a wide range of application requirements \cite{dia_review2}. As the key to promote the implementation of the whole speech processing system, speaker diarization has attracted the attention of a large number of researchers. However, in the face of complex acoustic scenes, a large number of speech overlaps and domain mismatches, the speaker diarization technology still faces great challenges \cite{wang2018speaker}.

The traditional speaker diarization method is mainly based on the system of module cascade, including voice activity detection, speech segmentation, speaker embedding extraction and clustering. Researchers often use i-vector\cite{i-vector} and x-vector\cite{x-vector} as the segment level embedding of speakers. In the clustering stage, K-means\cite{kmeans}, mean shift\cite{mean}, spectral clustering (SC)\cite{sc}, and agglomerative hierarchical clustering (AHC)\cite{AHC}\cite{AHC2} are often used to aggregate the regions of each speaker into separate clusters. Some systems have achieved good performance\cite{clu}. For example, the Bayesian HMM clustering of x-vector sequences (VBX) \cite{VBx} system performs two-stage clustering on x-vector to output more refined results. However, the clustering method is limited by its ability to assign a single speaker label to each speech segment, which makes it difficult to deal with overlapping speech segments.

Recently, researchers have made more and more research on end-to-end neural speaker diarization (EEND) systems\cite{e2e2019,e2e20192,e2e2022,e2etsvad}. This system regards the diarization task as a multi-label classification problem, that is, assigning multiple speaker probabilities to each frame of speech, so that the model can effectively deal with overlapping speech. The target speaker speech activity detection (TS-VAD) system \cite{ts-vad} proposed in recent years uses speech features and speaker embedding as input to directly predict the activity of each speaker in each frame, and has achieved good results. On this basis, researchers proposed the neural speaker diarization using memory-aware multi-speaker embedding (NSD-MA-MSE) \cite{MA-MSE}. The network uses a special memory module to extract clearer and more discriminative multi-speaker embedding from external data by the attention mechanism, thus replacing the low-quality i-vector representation in TS-VAD, achieving excellent results and becoming the champion system in the CHiME-7 challenge \cite{chime7,wan2023ustc}.

Although the audio-only speaker diarization system can achieve good results in many cases, in the actual scene, the audio often contains strong environmental noise, room reverberation, and multiple speakers overlap, resulting in the actual performance of the system less than expected \cite{gebru2017audio}. Unlike audio, visual information is not sensitive to ambient noise and reverberation, and can clearly determine which speaker is speaking according to the movement of the speaker's lips. Because of the strong correlation between voice information and visual information \cite{yehia1998quantitative}, especially the lip movement, researchers began to consider using the method of combining audio and video to solve the problem that the audio-only mode is difficult to solve. Audio-visual speaker diarization (AVSD), which combines audio and video information to get more accurate results, has become a research hotspot \cite{noulas2011multimodal,chung2019said,chung2020spot}. Recently, the end-to-end audio-visual speaker diarization method has received extensive attention \cite{he2022end,cheng2024multi}. By inputting audio features, video features, and speaker representation into the network, the speaker diarization results of each frame are directly predicted combined with multimodal information \cite{he2024quality}. This end-to-end architecture can not only make full use of the complementarity of multimodal information, but also significantly improve the robustness of the system in the complex acoustic environment through joint optimization, especially in low signal-to-noise ratio and multiple overlapping conversations.

Despite the significant progress made in speaker diarization based on deep learning, its practical application still faces the dual challenges of strong data dependence and insufficient model generalization \cite{wu2023semi}. From the perspective of data acquisition, building a high-quality speaker diarization dataset requires a lot of manpower and material resources. The annotation process not only requires professionals to accurately identify the speaker's identity, but also needs to precisely annotate the start and end time of speech activities. This fine-grained annotation work is often time-consuming and costly. Especially when dealing with complex scenarios such as multi-person interaction, speech overlap, and strong background noise, the annotation difficulty and cost are too high \cite{fu2021aishell,xu2022ava}.
This high annotation cost seriously constrains the construction of large-scale datasets, which in turn limits the performance improvement of deep learning models. The emergence of semi-supervised learning has led researchers to start thinking about how to improve model performance without relying on a large amount of annotated data. However, the existing semi-supervised speaker diarization methods based on pseudo-labels are limited to generating pseudo-labels for unannotated data using speaker diarization models pre-trained on out-of-domain data \cite{takashima2021semi}. However, when there are significant differences between datasets, there is a high probability of encountering a domain mismatch problem, resulting in poor quality of pseudo-labels. The method \cite{wu2023semi} avoids this problem but still requires some labeled data to train the model, which makes its applicability not very strong.

From the perspective of model generalization ability, existing methods generally have the problem of poor adaptability to different scenarios. This is mainly due to two limitations: Firstly, most models are trained and optimized only on datasets in specific fields, resulting in their learned feature representations having a strong dependence on the scenario. For example, models trained on meeting data may have difficulty adapting to the needs of different scenarios, such as telephone customer service or outdoor communication. Secondly, most of the existing public datasets are limited to a single scenario, lacking diversity across scenarios and domains \cite{watanabe2020chime,wang2023multimodal,vinnikov2024notsofar}. They are not very applicable to complex scenarios, which hinders the wide application and further development of speaker diarization technology. Currently, there is relatively little research on this problem.

Accordingly, in this study, to solve the above two problems, we proposed a semi-supervised speaker diarization dataset construction method guided by audio and video, as well as a fine-tuning method for scene-related speaker diarization models. We also open-sourced a multi-modal, multi-scenario, and multi-language speaker diarization (M3SD) dataset and the code for building the dataset to support in-depth research in this field. First of all, we proposed a method for constructing the speaker diarization dataset guided by audio and video. In actual scenarios, the audio data may be difficult to use due to factors such as noise, and the audio-only speaker diarization methods are sensitive to changes in the scene, resulting in poor quality of the generated pseudo-labels. However, the audio-visual speaker diarization method can avoid similar problems. Specifically, we build an automated speaker diarization data collection and annotation system to generate pseudo-labels for massive unlabeled data. The system covers modules such as data collection and cleaning, data preprocessing, and pseudo-label generation based on multiple speaker diarization models. Using this technology, we have built a large-scale speaker diarization dataset, covering interviews, online/offline meetings, speeches, movies, debates, and daily conversations, and including Chinese, English, Japanese, and other languages, etc. The dataset contains 1,372 records, 770+ hours of data, and a large number of different speakers. This design can fill the gap of insufficient data scenarios covered by existing datasets, making trained models have strong generalization capabilities.

In summary, our main contributions are as follows: We proposed a method for constructing a multi-modal, multi-scenario and multi-language speaker diarization dataset guided by audio and video, and released a large-scale speaker diarization dataset based on pseudo-labels. Section II introduces the relevant work. Section III introduces the construction method of the dataset and gives an overall introduction to the dataset.

\section{Related work}

With the development of speaker diarization, researchers have released some related datasets, which include different languages and target scenarios, and have contributed to the development of speaker diarization tasks. In the early days, many audio-only datasets were released. CALLHOME \cite{martin2001nist} is one of the most widely used datasets in early research, containing more than 500 telephone recordings and covering different languages and the number of speakers; CHiME-6 \cite{watanabe2020chime} dataset is mainly aimed at the dinner conversation scene of 4 people, where speakers can move freely in different rooms, with highly free conversations; DIHARD \cite{ryant2018first,ryant2019second,ryant2020third} series datasets are mainly aimed at the generalization ability of speaker diarizationsystems in multiple scenarios, and the datasets include high-difficulty scenarios such as restaurants and online videos; Alimeeting dataset is based on the M2MeT \cite{M2Met} chellenge and is aimed at the Chinese conference scene.

With the development of multimodal speaker diarization, audio-visual speaker diarization datasets have also been gradually released. AMI \cite{kraaij2005ami} is the earliest audio-visual speaker diarization dataset, targeting English conference scenarios, with each conference containing 3 to 5 speakers; AVA-AVD \cite{xu2022ava} dataset is 29 hours of annotating movie clips and has a certain degree of diversity; MISP \cite{wang2023multimodal,gao2025multimodal} series datasets target real Chinese home scenarios, which are quite challenging; MSDWILD \cite{liu2022msdwild} dataset is 80 hours of vlog videos crawled from video websites and manually annotated.

Despite the existence of the above-mentioned datasets in the field of speaker diarization, there are still some problems. Firstly, compared with fields such as speech recognition and natural language processing, there is still a significant gap in the scale of data resources. Secondly, most of the data still focuses on simple scenarios such as meetings. Although speaker diarization technology is most widely used in meeting scenarios, there are still some difficult application scenarios. Due to the poor generalization of deep learning models, which limits the performance of the system. Although the DIHARD series of data includes data from multiple scenarios, the amount of data for each scenario is relatively small, and it only has a single audio modality, which can support fewer research. Moreover, most datasets still only contain English, a small part includes Chinese, and there is even less support for other minor languages. Therefore, the existing datasets are not enough for the future development of speaker diarization technology, and we need to build larger-scale data resources that include more scenarios and more languages.

\section{M3SD Dataset}
We have proposed a Multi-modal, Multi-scenario, and Multi-language Speaker Diarization Dataset (M3SD). This dataset contains 770+ hours and 1372 segments of conversations, including many scenarios such as online/offline meetings, debates, speeches, home/outdoor conversations, movies, and news broadcasts. It also includes multiple languages. The dataset also includes video metadata to support multimodal research. Additionally, we have open-sourced the code for data collection to facilitate the construction of larger datasets. This process is fully automated, ensuring data quality through rigorous data cleaning and preprocessing, and generating pseudo-labels using pre-trained audio-only as well as audio-visual speaker diarization models, without the need for manual annotation. In this section, we will introduce the overall data collection process.

The method proposed in this paper for constructing the speaker diarization dataset guided by audio and video is an automated semi-supervised method, which is carried out through the framework of data collection - data processing - generating pseudo labels, and iterative training. Figure 1 shows the overall method of constructing the dataset. Firstly, according to the carefully designed search terms, we crawl video data on the video websites YouTube and BiliBili. Secondly, through data cleaning, we remove poor-quality data (such as low-quality speech, low-resolution videos, and audio-visual synchronization issues) and retain valid data that meets our needs. This step mainly includes modules such as speech quality assessment, video quality assessment, and audio-visual synchronization detection. After that, we preprocess the audio and video data, which mainly includes face detection, face trajectory tracking, and extraction of the lip region of interest (ROI). Then, we input the data into the pre-trained audio-only speaker diarization model and the audio-visual speaker diarization model respectively. Finally, we use voting fusion to integrate the output results of the two models to obtain the speaker diarization pseudo-labels. The reason for using two models is that the quality of video data on the Internet is uneven. Although we have already gone through a rigorous data cleaning process, there may still be cases where the audio or video quality is too poor in a certain segment, which may affect the quality of the generated pseudo-labels.
Although the robustness of audio-visual speaker diarization is relatively strong, and in most cases, the effect is better than the audio-only speaker diarization method, but when the video quality is too poor, such as long-time occlusion, off-screen voice, and blurry camera shake, etc., it may also drag down the final result. Therefore, it is necessary to combine the results of the two methods to get a better effect. After the above steps, we can obtain the pseudo-labels of the dataset, and then we can use the pseudo-labels to iteratively retrain or fine-tune the speaker diarization model, so that the model performance and the quality of pseudo-labels can get a virtuous circle, and further improve the data quality. Next, we will introduce the specific implementation of each module in detail, in order.

\begin{figure}[!t]
\centering
\includegraphics[width=0.8\linewidth]{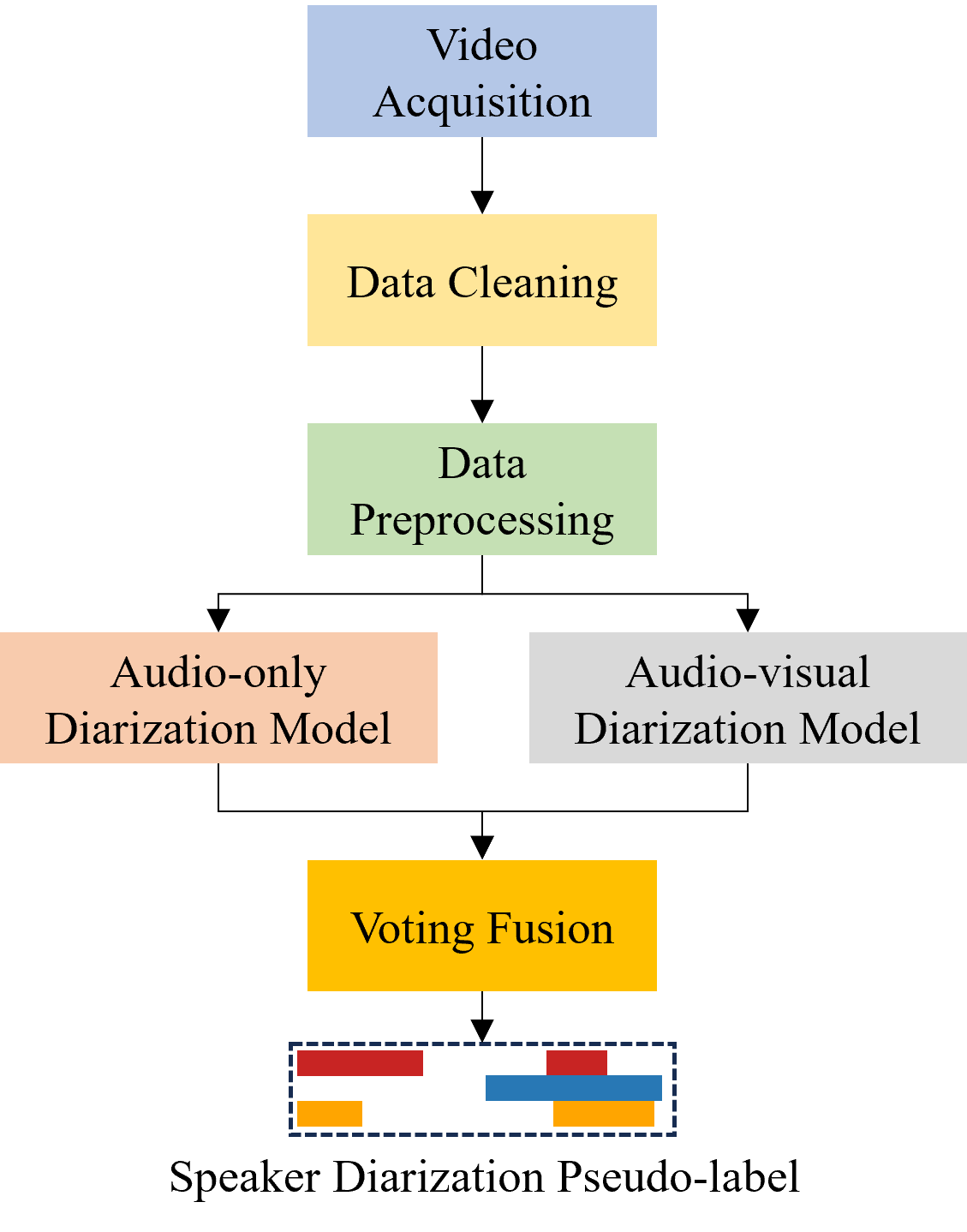}
\caption{Flowchart of dataset construction process.}
\label{fig_1}
\end{figure}

\subsection{Video Acquisition}

In order to build a diverse speaker diarization dataset, we need to crawl video data on video websites such as YouTube and BiliBili according to keywords of different themes. We first design a set of keyword lists, and search through different combinations of keywords, such as conference in Chinese, speech in English, debate in Japanese, etc. And we search these keywords in different languages by changing the website location or using translation software. The reason for this design is that we need to obtain as much video data in different scenes and different languages as possible, and expand the diversity of the dataset, so as to improve the generalization ability of subsequent training models. Moreover, when crawling videos, we directly filter out those that are less than three minutes long and set the video resolution to 720p to ensure the quality of the generated pseudo-labels. The examples of the content covered by the dataset are shown in Figure 2. Compared with previous datasets, the complexity and diversity of the scenarios in the dataset proposed in this paper are a major highlight. It basically covers most of the application scenarios that can be involved in the field of speaker diarization, such as online/offline meetings, interview talks, speeches, news broadcasts, multi-person debates, outdoor conversations, etc.
The dataset covers a more comprehensive range of scenarios, has a larger scale of data, and supports the audio-visual research, which gives it some distinct advantages.

\begin{figure}[!t]
\centering
\includegraphics[width=1\linewidth]{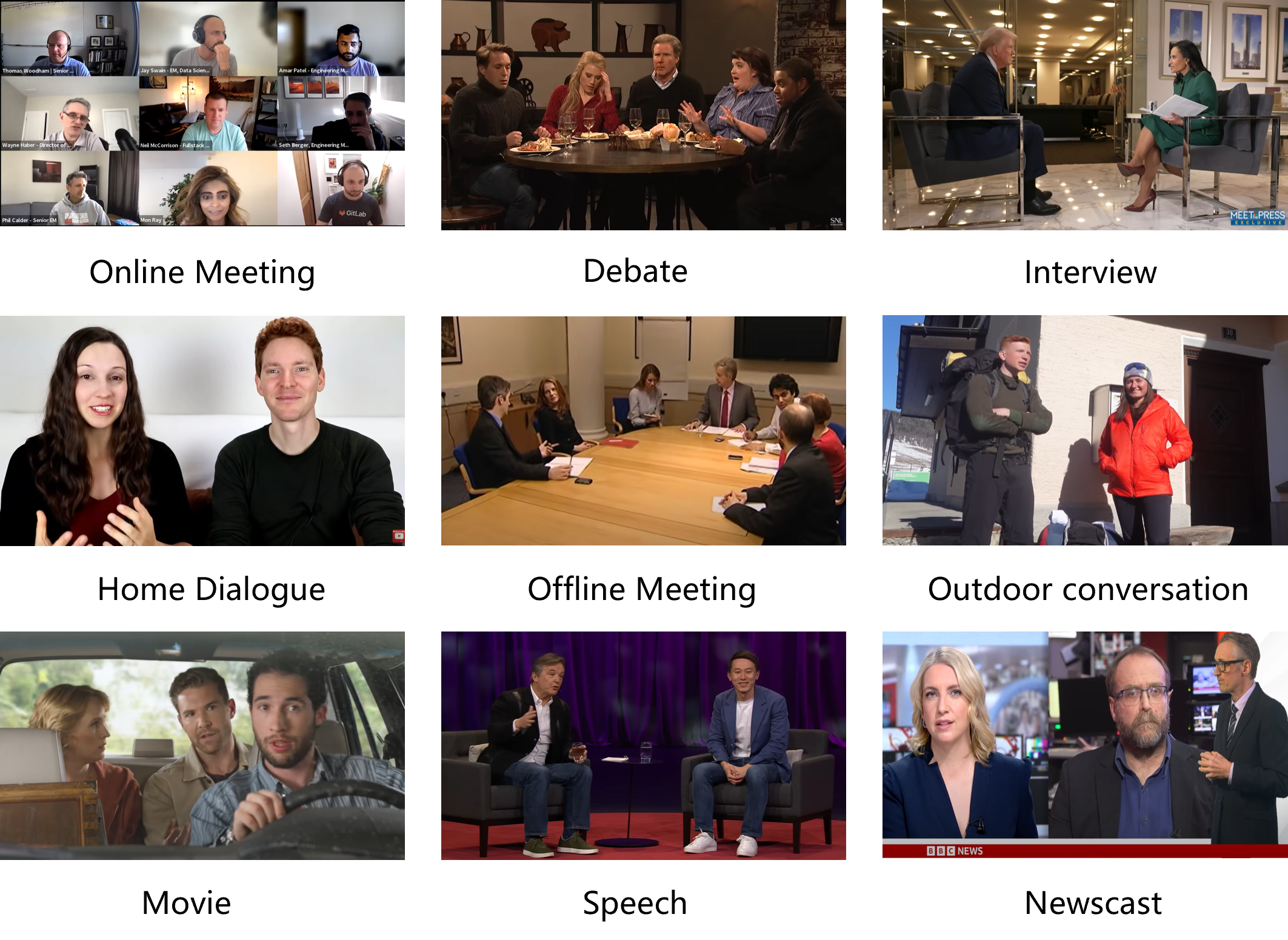}
\caption{Example of content covered by the dataset.}
\label{fig_2}
\end{figure}

\subsection{Data Cleaning}

Due to the uneven quality of data on the Internet, there may be problems such as poor voice quality, blurry and shaky videos, audio and video desynchronization, and long-time off-screen voice, which seriously affect the quality of subsequent pseudo-label generation. It is inevitable to spend a lot of time and effort on careful manual screening. Therefore, we need to design a series of automated data cleaning processes to discard poor-quality data and retain high-quality data.
As shown in the upper part of Figure 3. The data cleaning process in this article mainly consists of several steps, including shot detection and segmentation, audio/video extraction, speech quality evaluation, video quality evaluation, and audio-video synchronization detection. Through this process, we can conduct rigorous data cleaning on the original data. Next, we will introduce in detail the specific implementation of each module.

\textbf{Shot detection:}
Videos on the Internet often have a lot of post-production editing, such as adding opening and closing credits, switching between multiple camera angles, and splicing video clips, which can lead to the discontinuity of video scenes and mismatch with real scenes. This will affect subsequent processing, such as causing face detection and tracking to fail, and leading to changes in the speaker's ID and number, etc. Therefore, the first thing to do is to perform shot detection and segmentation on the video, and cut the original video at the time points where the scene changes significantly, to obtain multiple video clips in the same scene without rapid scene changes.
The shot detection method we use is the PySceneDetect toolkit \cite{bieda2021approach}. By calculating the difference value of each pixel between two consecutive frames, we generate a differential image. The system sets a threshold. When the cumulative degree of pixel changes in the differential image exceeds this threshold, it is determined that there has been a transition in the scene. The threshold set in this paper is 30, which can achieve good detection results while avoiding misjudgment. After performing shot detection, we then cut the video according to the change points.

\textbf{Audio/Video Extraction:}
After cutting the video, we need to extract the audio from the video so that we can process them separately later. In this article, we use the FFmpeg toolkit to extract it, and get the audio with a sampling rate of 44k and the video with a resolution of 720p. High-quality audio and video will help us carry out subsequent tasks.

\begin{figure}[!t]
\centering
\includegraphics[width=0.8\linewidth]{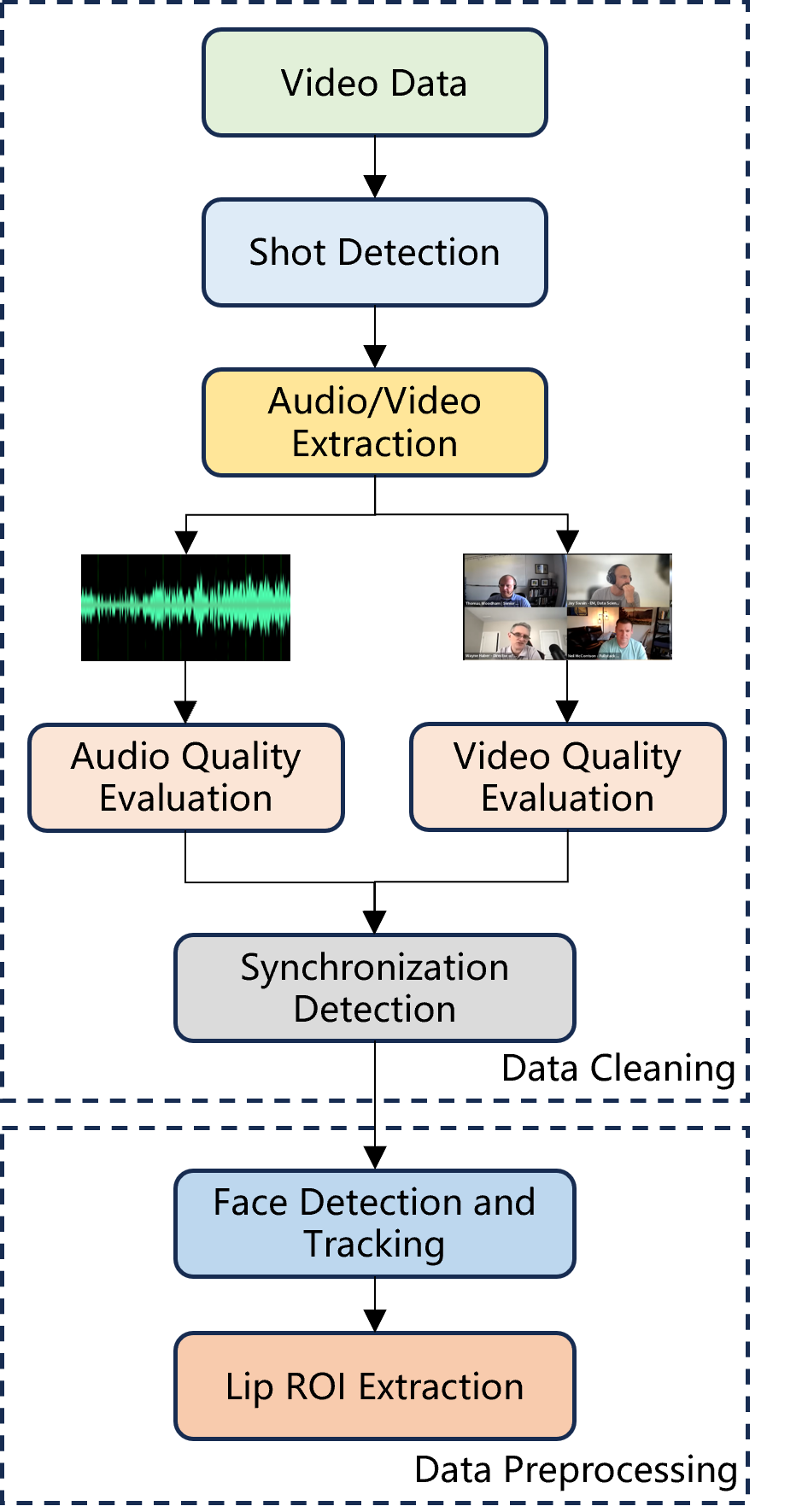}
\caption{Data cleaning and data preprocessing flowchart.}
\label{fig_3}
\end{figure}

\textbf{Audio Quality Evaluation:}
Since the speaker diarization task relies on high-quality audio and video, and the quality of online videos varies greatly, it is necessary to carry out a quality evaluation of audio and video. Since online data does not have reference speech, we cannot judge the speech quality by evaluating the difference between the output speech and the original speech like the speech enhancement task. Therefore, we need to carry out a reference-free speech quality evaluation.
Deep Noise Suppression Mean Opinion Score (DNSMOS) \cite{reddy2022dnsmos} can evaluate speech quality without reference speech, and it is highly correlated with human subjective evaluation. As a non-intrusive perceptual objective speech quality metric, this method adopts the ITU-T P.835 standard to predict human perception scores of speech quality, generating three key scores: speech quality (SIG), background noise quality (BAK), and overall audio quality (OVRL), with a score of 0-5 (5 being the best).
Due to its good generalization ability in various noise types and environments, it is very suitable for evaluating the speech quality of network videos. In order to make the subsequent speech be well used in the speaker diarization task, this paper sets the score threshold to 3 points.

\textbf{Video Quality Evaluation:}
Similar to audio, we also need to conduct a reference-free quality assessment of videos to eliminate low-quality data such as blurry shots, jitter, and distortion. By comparing the scores output by various methods with real human perception, we ultimately chose the Multi-Dimensional Video Quality Assessment (MD-VQA) \cite{zhang2023md} algorithm.
This method is a multi-dimensional, reference-free video quality assessment method specifically for network videos or live videos. It can integrate semantic, distortion, motion, and other multi-dimensional information of the video, and perform temporal and spatial fusion to measure the absolute quality of the video. Compared with traditional methods, this multi-dimensional evaluation is closer to human perception of video quality, and has higher accuracy and interpretability. By setting a certain score threshold (60 in this paper), we can eliminate low-quality videos to ensure the quality of generating pseudo-labels.

\textbf{Audio-visual Synchronization Detection:}
Videos crawled from the Internet may have audio and video that are out of sync, or all be dubbed, narrated, or other off-screen voices, which will lead to incorrect results in the audio-visual speaker diarization. Therefore, it is necessary to determine whether the audio and video are synchronized, and whether there is more than one face synchronized with the audio.
SyncNet \cite{chung2017out} is a deep learning model specifically designed to detect audio-visual synchronization. It determines whether the audio and video are synchronized by analyzing the audio signal and video frames in the video. The unique feature of this method is that it can handle "wild" scenes, that is, videos containing various complex conditions such as different lighting, shooting angles, and background noise. The network ultimately outputs the audio-visual offset and confidence score for each face track. In this paper, an offset less than or equal to 5 and a confidence score greater than or equal to 1 are considered to be audio-visual synchronization.
In addition to detecting whether the audio and video are synchronized, this step can also check whether the lip movement trajectory of one or more speakers in the video segment is highly correlated with the speech, thus filtering out videos without any speakers. After going through the above rigorous data cleaning process, we can obtain data resources of relatively high audio and video quality, thereby ensuring the accuracy of subsequent pseudo-label generation.

\subsection{Data Preprocessing}

After the data cleaning, we need to preprocess the data to meet the input requirements of subsequent tasks. As shown in the lower half of Figure 3, preprocessing mainly targets the video part, including face detection, face trajectory tracking, and lip ROI extraction. Accurate face trajectories and high-quality lip regions will be key factors affecting the accuracy of the subsequent audio-visual speaker diarization module.

\textbf{Face Detection}
Since the audio-visual speaker diarization system requires high-precision lip input, we choose to first perform face recognition, and then perform lip recognition and extraction within the face area. This approach can avoid the accuracy problems brought about by direct lip extraction. We chose the RetinaFace \cite{deng2020retinaface} face detection algorithm. In network videos, the most important feature is the different sizes of faces, so it is necessary to have strong detection capabilities for faces of different sizes. Due to its unique design, the RetinaFace algorithm can handle this situation very well and achieve excellent results. Its core goal is not only to detect the location of a person's face, but also to extract key points of the face (such as the eyes, nose, and corners of the mouth) and pose information (pitch angle, yaw angle), which is suitable for scenarios such as online videos where the conditions are not fixed.

\begin{figure*}[!t]
\centering
\includegraphics[width=0.85\linewidth]{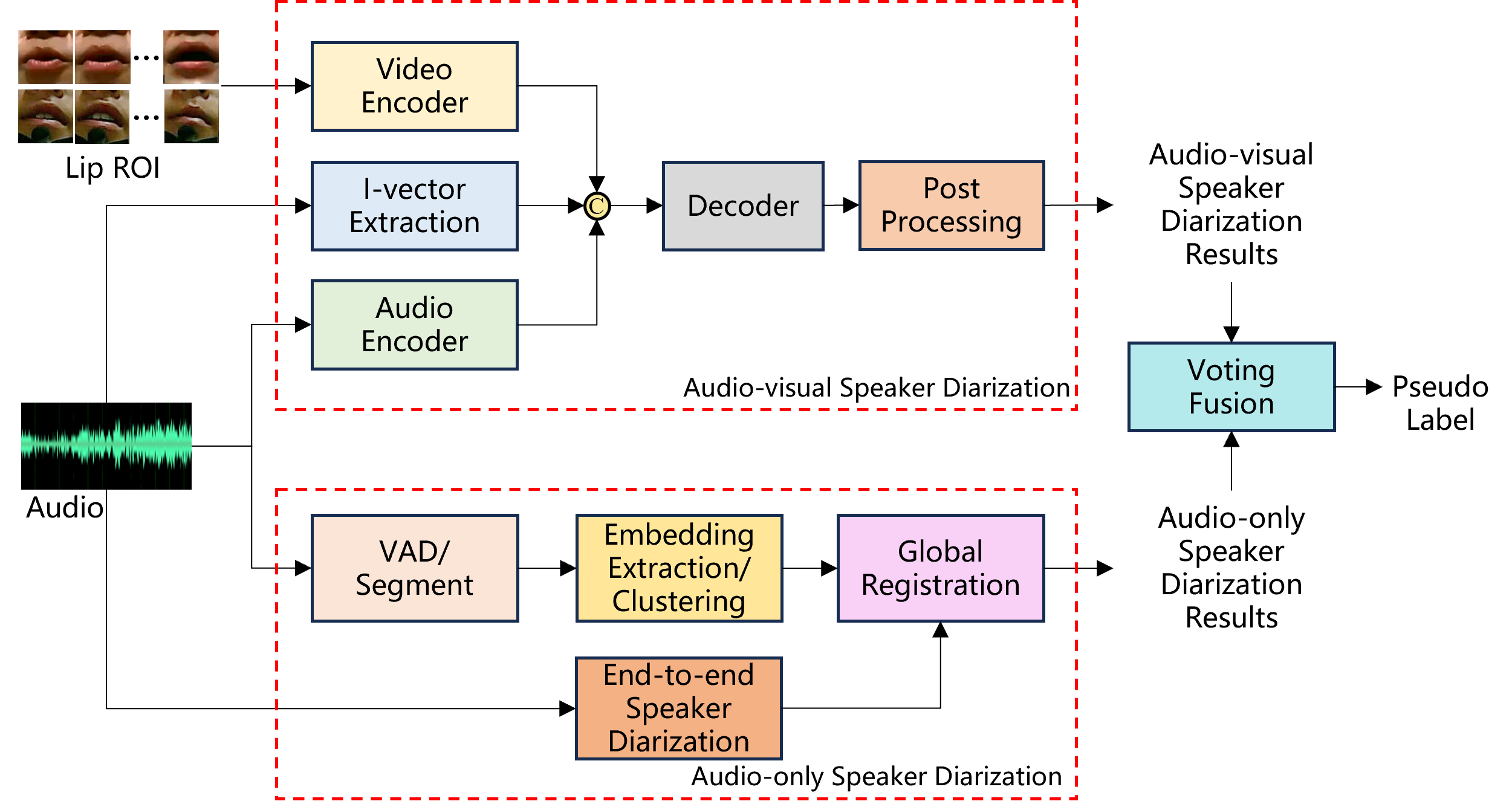}
\caption{Pseudo-label generation based on the pre-trained speaker diarization model.}
\label{fig_4}
\end{figure*}

\textbf{Face Tracking}
After obtaining the face detection results, we need to track the face trajectory to assign a speaker ID. In this paper, we choose the DeepSORT \cite{wojke2017simple} algorithm, which updates the trajectory status of the target in real time in each frame of the video and assigns a trajectory ID. It combines motion information (predicting the target location through the Kalman filter) and appearance information (describing the target characteristics through deep features), thus achieving efficient and accurate face trajectory tracking. Compared with traditional tracking algorithms, DeepSORT significantly improves the tracking performance and robustness in complex scenarios through the appearance features extracted by deep learning.

\textbf{Lip ROI Extraction}
The input of the audio-visual speaker diarization is audio and the region of interest (ROI) of the lips, so it is also necessary to extract the speaker's lip area. In the previous steps, we have obtained the face detection box of each speaker, and then we can detect the key points of the lips within the face box, thereby extracting the lip area. We choose the MediaPipe \cite{lugaresi2019mediapipe} framework for key point detection. This method effectively overcomes the detection errors caused by distance changes or viewing angle shifts, and is more suitable for large-scale network videos.
In face key point extraction, MediaPipe can detect up to 468 key points, including many points around the lips, which can provide very detailed information about the shape and contour of the lips. If some key points are obscured, we can determine the location of the area based on other key points.
The extracted key points are [61, 185, 40, 39, 37, 0, 267, 269, 270, 409, 291, 146, 91, 181, 84, 17, 314, 405, 321, 375, 291, 57, 430, 164, 287, 200, 210], covering the entire lip area and the surrounding area.

\subsection{Pseudo-label Generation}

After obtaining the audio and lip ROI, we input the data in parallel into the pre-trained audio-only speaker diarization model and the audio-visual speaker diarization model. Finally, we use the voting fusion method to integrate the prediction results of the two models to generate speaker diarization pseudo-labels. The reason for considering audio-only speaker diarization in addition to audio-visual speaker diarization is that although we have already gone through a rigorous data cleaning process, the audio or video quality of some segments may still be poor, which may affect the generation effect of pseudo-labels. Generally, the performance of the audio-visual speaker diarization model is better than that of the audio-only model, and its robustness is stronger.
However, when the video quality is poor, such as long-time occlusion, off-screen voice, or blurry shots due to camera shake, its performance may be limited to some extent. If the audio quality of the clip is good, then the effect of an audio-only speaker diarization may exceed that of an audio-visual speaker diarization. Therefore, in order to improve the accuracy and stability of overall detection, it is necessary to comprehensively utilize the output results of the two models to obtain a better pseudo-label generation effect. Figure 4 shows this pseudo-label generation method based on the pre-trained speaker diarization model.

\textbf{Audio-visual Speaker Diarization:}
The end-to-end audio-visual speaker diarization network we use originates from research \cite{he2022end}, as shown in the upper part of Figure 4. The input of this network consists of the original audio and the lip ROI. By highly integrating audio and video features, we obtain highly robust and accurate speaker diarization results. First of all, this paper uses AMI, MISP, MSDWild, AVA-AVD, and other audio-visual speaker diarization datasets to pre-train the model. These datasets provide rich and diverse audio and video data, covering different scenarios and speaker combinations.
Through pre-training on these datasets, the model can learn a wide range of speaker features and scene feature expressions, thus having good generalization ability and robustness, which is suitable for dealing with network videos, which are complex data in multiple scenarios. This method integrates multi-modal information of audio, video, and speaker representation, and end-to-end predicts the probability of the speaker's speech activity in each frame. This feature-level integration makes full use of the correlation and complementary advantages of multi-modal information, and has strong robustness.

\textbf{Audio-only Speaker Diarization:}
To alleviate the problem of inaccurate audio-visual speaker diarization resulting from the poor quality of some videos, we further combine the audio-only speaker diarization for processing. We experimented and compared various audio-only speaker diarization methods, and finally chose the 3D-Speaker multi-speaker diarization network \cite{chen20253d}.
The traditional module cascade speaker diarization network has good generalization, but since it can only assign one speaker to each segment of speech, it cannot handle overlapping segments of speech. The end-to-end speaker diarization can handle overlapping areas, but it is also limited by poor generalization, difficulty in estimating the number of speakers, and difficulty in processing longer audio etc. Therefore, by combining the advantages of the two systems, we can further improve the accuracy and robustness of the speaker diarization task in complex acoustic scenarios.
As shown in the lower part of Figure 4, the implementation of this method is divided into two main branches: one is based on feature extraction and unsupervised clustering of speaker diarization at the module level, which is used for global speaker number detection and coarse-grained segment-level result output; the other is based on end-to-end speaker diarization, responsible for overlap segment detection, which is used for fine-grained frame-level result output. Finally, the results of the two are integrated through the global alignment module. This combination fully utilizes the global consistency advantage of traditional clustering and the overlap segment processing capability of deep learning models, thereby achieving a more robust speaker diarization effect.

\textbf{Voting Fusion:}
We use the DOVER-LAP \cite{raj2021dover} to fuse the results of the audio-visual speaker diarization and the audio-only speaker diarization, in order to obtain more accurate pseudo-labels.

In summary, this is our proposed method for building a speaker diarization dataset, which can make pseudo-labels more accurate through iterative training. So far, we have built and open-sourced a dataset containing more than 770 hours of data, covering multi-modal, multi-scenario, and multi-language, which can provide strong support for training a deep learning-based speaker diary network.

\bibliographystyle{IEEEtran}
\bibliography{reference.bib}

\vfill

\end{document}